\documentclass[12pt]{article}
\usepackage{epsfig}

\font\blackboard=msbm10 at 12pt
\font\blackboards=msbm7
\font\blackboardss=msbm5
\newfam\black
\textfont\black=\blackboard
\scriptfont\black=\blackboards
\scriptscriptfont\black=\blackboardss

\newcommand{\junk}[1]{}

\newcommand{\ba}{\begin{array}}
\newcommand{\ea}{\end{array}}
\newcommand{\be}{\begin{equation}}
\newcommand{\ee}{\end{equation}}
\newcommand{\bea}{\begin{eqnarray}}
\newcommand{\eea}{\end{eqnarray}}
\newcommand{\beas}{\begin{eqnarray*}}
\newcommand{\eeas}{\end{eqnarray*}}

\def\laplace{{\kern1pt\vbox{\hrule height 1.2pt\hbox{\vrule width
1.2pt\hskip
  3pt\vbox{\vskip 6pt}\hskip 3pt\vrule width 0.6pt}\hrule height
  0.6pt}
  \kern1pt}}
\def\scriptlap{{\kern1pt\vbox{\hrule height 0.8pt\hbox{\vrule width
  0.8pt
  \hskip2pt\vbox{\vskip 4pt}\hskip 2pt\vrule width 0.4pt}\hrule height
  0.4pt}
  \kern1pt}}

\def\roughly#1{\raise.3ex\hbox{$#1$\kern-
.75em\lower1ex\hbox{$\sim$}}}

\newcommand{\NP}{{\em Nucl.\ Phys.\ }}

\newcommand{\PL}{{\em Phys.\ Lett.\ }}
\newcommand{\PR}{{\em Phys.\ Rev.\ }}

\newcommand{\gone}[1]{}
\begin{document}
\pagestyle{plain}
\setcounter{page}{1}

\baselineskip16pt

\begin{titlepage}

\begin{flushright}
MIT-CTP-3096\\
NSF-ITP-01-17\\
hep-th/0103085
\end{flushright}
\vspace{13 mm}

\begin{center}

{\Large \bf Open string field theory without open strings}

\end{center}

\vspace{7 mm}

\begin{center}

Ian Ellwood and Washington Taylor\footnote{Current
address:
Institute for Theoretical Physics,
University of California,
Santa Barbara, CA 93106-4030; {\tt  wati@itp.ucsb.edu}}

\vspace{3mm}
{\small \sl Center for Theoretical Physics} \\
{\small \sl MIT, Bldg.  6} \\
{\small \sl Cambridge, MA 02139, U.S.A.} \\
{\small \tt iellwood@mit.edu, wati@mit.edu}\\
\end{center}

\vspace{8 mm}

\begin{abstract} Witten's cubic open string field theory is expanded
around the perturbatively stable vacuum, including all scalar
fields at levels 0, 2, 4 and 6.  The (approximate) BRST cohomology of
the theory is computed, giving strong evidence for the absence of
physical open string states in this vacuum.  \end{abstract}

\vspace{1cm}
\begin{flushleft}
March 2000
\end{flushleft}
\end{titlepage}
\newpage

\section{Introduction}

The 26-dimensional open bosonic string has a tachyon in its spectrum
with $M^2 = -1/\alpha'$.  The presence of this tachyon indicates that
the perturbative vacuum of the theory is unstable.  While some early
work \cite{ks-open} indicated the
possible existence of a more stable vacuum at lower energy (see also
\cite{Bardakci-Halpern-tachyon,Bardakci-tachyon}), until
fairly recently the significance of this other vacuum was not
understood, and the tachyon was taken to be an indication of
fundamental problems with the open bosonic string.

In 1999, Sen suggested that the open bosonic string should be
interpreted as ending on an unstable space-filling
D25-brane \cite{Sen-universality}.  Sen argued that the condensation of
the tachyon should correspond to the decay of the
D25-brane, and that it should be possible to give an analytic
description of this condensation process using the language of
Witten's cubic open string field theory \cite{Witten-SFT}.  In
particular, Sen made three concrete conjectures:

\begin{enumerate}
\item[a)] The difference in the action between the unstable vacuum and
the perturbatively stable vacuum should be $\Delta E =VT_{25}$, where
$V$ is the volume of space-time and $T_{25}$ is the
tension of the D25-brane.
\item[b)] Lower-dimensional D$p$-branes should be realized as soliton
configurations of the tachyon and other string fields.
\item[c)] The perturbatively stable vacuum should correspond to the
closed string vacuum.  In particular, there should be no physical open
string excitations around this vacuum.
\end{enumerate}

Conjecture (a) has been verified to a high degree of precision in
level-truncated cubic open string field theory
\cite{Sen-Zwiebach,Moeller-Taylor} and has been shown exactly using
background independent string field theory
\cite{Gerasimov-Shatashvili,kmm,Ghoshal-Sen}.  Conjecture (b) has been
verified for a wide range of single and multiple D$p$-brane
configurations, using both the cubic and background independent
formulations of SFT (see \cite{Harvey-Komaba} for a review and further
references).  To date, however, little concrete evidence has been put
forth either for the decoupling of open strings in the perturbatively
stable vacuum or for the interpretation of this state as the closed
string vacuum.  In this note we explicitly compute the scalar open string
spectrum in the cubic open string field theory expanded around the
perturbatively stable vacuum, using the level-truncation
approximation.  
The results of this computation give strong evidence
that there are no physical open string states in this vacuum, and that
the open strings are removed from the spectrum by purely classical
effects in the string field theory.

\section{String Field Theory in the Stable Vacuum}

We begin with a brief summary of Witten's cubic formulation of open
bosonic string field theory \cite{Witten-SFT} (see
\cite{lpp,Gaberdiel-Zwiebach} for reviews).  The string field
$\Phi$ contains an infinite family of space-time fields, one field
being associated with each state in the open string Fock space.
Physical fields are associated with states in the Hilbert space of
ghost number one.  The string field may be formally written as
\begin{equation}
\Phi = \phi (p)| \hat{0}; p \rangle+
A_\mu (p) \alpha^\mu_{-1}| \hat{0}; p \rangle + \cdots
\end{equation}
where $| \hat{0} \rangle$ is the ghost number one vacuum related to
the $SL(2,R)$-invariant vacuum $| 0 \rangle$ through $| \hat{0} \rangle=
c_1| 0 \rangle$.   
Witten's cubic string field theory action is
\begin{equation}
S = - \frac{1}{2}  \int \Phi \star Q \Phi - \frac{g}{3}  \int \Phi \star
\Phi \star \Phi
\label{eq:SFT-action}
\end{equation}
where $Q$ is the BRST operator of the string theory and the ``star
product'' $\star$ is defined by dividing each string evenly into
two halves and ``gluing'' the right side of one string to the left side
of the other through a delta function interaction.  The action
(\ref{eq:SFT-action}) is invariant under the stringy gauge
transformations
\begin{equation}
\delta \Phi = Q \Lambda + g \left( \Phi\star \Lambda -
\Lambda\star \Phi \right)
\label{eq:gauge}
\end{equation}
where $\Lambda$ is a ghost number zero string field.

While there are an infinite number of component fields in the string
field $\Phi$, for any  particular component fields
the quadratic and cubic interactions in
(\ref{eq:SFT-action}) and the related terms in the gauge
transformations (\ref{eq:gauge})
can be
computed in a straightforward fashion using a Fock space
representation of the BRST operator $Q$ and the star product
\cite{Gross-Jevicki-12,cst,Samuel}.
It has been found \cite{ks-open,WT-SFT} that truncating the theory by
including only fields up to a fixed level $L$ is an effective
approximation technique for many questions relevant to the tachyon
condensation problem.  (By convention the tachyon is taken to have
level zero).  At fixed level $L$, the theory can be further simplified
by only considering interactions between fields whose levels total to
some number $I < 3L$.  Empirical evidence
\cite{ks-open,Moeller-Taylor} indicates that truncating at level $(L,
I) = (L, 2L)$ is the most effective cutoff to maximize accuracy for a
fixed number of computations and that calculations at truncation
level $(L, 2L)$ give similar results to calculations at truncation
level $(L, 3L)$.

Sen's conjecture states that there is a Lorentz invariant
solution $\Phi_0$ of the full string field theory equations of motion
\begin{equation}
Q \Phi_0 = -g \Phi_0 \star \Phi_0\,
\label{eq:SFT-EOM}
\end{equation}
corresponding to the closed string vacuum without a D25-brane.  The
existence of a nontrivial solution to (\ref{eq:SFT-EOM}) has been
analyzed in the level-truncated theory
\cite{ks-open,Sen-Zwiebach,Moeller-Taylor}.  In the level (0, 0)
truncation, the tachyon potential is simply $-\frac{1}{2}\phi^2 +g
\kappa \phi^3$, where $\kappa$ is a numerical constant.  This
potential gives a locally stable vacuum at $\phi_0 = 1/(3g \kappa)$.
Evaluating the potential at this point gives 68\% of Sen's conjectured
value $T_{25}$ for the energy gap between the unstable vacuum and the
perturbatively stable vacuum.  When other scalar fields are included
in the string field by raising the level at which the theory is
truncated, many of these fields couple to the tachyon $\phi$ and take
expectation values when $\phi$ becomes nonzero, but similar solutions
to the level-truncated string field theory equations of motion
continue to exist.  In the level (4, 8) truncation, the energy gap
between the two vacua becomes $98.6\%$ of the predicted value
\cite{Sen-Zwiebach}, and in the level (10, 20) truncation the energy
gap becomes $99.91\%$ of the predicted value \cite{Moeller-Taylor}.
As the level of truncation is increased, the vacuum expectation values
of the scalar fields converge rapidly, so that the level (10, 20)
values for the field values in the vacuum appear to be within less
than $1\%$ of their exact values for low-level fields.  These results
provide us with a close approximation to a field $\Phi_0$ satisfying
(\ref{eq:SFT-EOM}), which we can use to study the perturbatively
stable vacuum in the level truncated theory.

We can describe the physics around a nontrivial vacuum $\langle\Phi
\rangle = \Phi_0$ satisfying the equation (\ref{eq:SFT-EOM}) by
shifting the string field
\begin{equation}
\Phi = \Phi_0 + \tilde{\Phi}\,.
\end{equation}
In terms of the new field $\tilde{\Phi}$, the action becomes 
\begin{equation}
S = S_0 - \frac{1}{2}  \int \tilde{\Phi} \star \tilde{Q} \tilde{\Phi} -
\frac{g}{3}  \int \tilde{\Phi} \star 
\tilde{\Phi} \star \tilde{\Phi}
\label{eq:SFT-t-action}
\end{equation}
where the new BRST operator $\tilde{Q}$ acts on a string field $\Psi$ of
ghost number $n$ through
\begin{equation}
\tilde{Q} \Psi= Q \Psi +  g \left(
\Phi_0 \star \Psi - (-1)^n \Psi \star \Phi_0 \right)\,.
\label{eq:tq}
\end{equation}
The identity
\begin{equation}
\tilde{Q}^2 = 0
\label{eq:q2}
\end{equation}
for the new BRST operator follows from (\ref{eq:SFT-EOM}).
The BRST invariance of the level-truncated approximation to the vacuum
$\Phi_0$ was studied in \cite{Hata-Shinohara}

We are interested in studying the physics of the new string field
theory defined through (\ref{eq:SFT-t-action},\ref{eq:tq}).  According
to Sen, the vacuum $\Phi_0$ should be the closed string vacuum, and
should not admit open string excitations.  To study the spectrum of
excitations of the theory, we need to explicitly calculate the
quadratic terms in the action, or equivalently to compute the action
of the new BRST operator (\ref{eq:tq}) on a general string field.
This requires us to compute all cubic couplings in the original string
field theory of the form
\begin{equation}
t_{ijk} (p) \;\phi_i (0) \phi_j (-p) \phi_k (p)\, .
\label{eq:0pp-form}
\end{equation}
In this letter we restrict attention to scalar excitations, so we need
to compute all terms of the form (\ref{eq:0pp-form}) where $\phi_i,
\phi_j$ and $\phi_k$ are scalar fields.  Because $\phi_j, \phi_k$ are
momentum-dependent, we must include among these fields longitudinal
polarizations of all higher-spin tensor fields as well as the
zero-momentum scalars $\phi_i (0)$ which can take nonzero vacuum
expectation values in $\Phi_0$.  We restrict attention in this letter
to scalars at even levels, which decouple from odd-level scalars at
quadratic order due to the twist symmetry of the theory
\cite{ks-open,Sen-Zwiebach}.

With the assistance of the symbolic manipulation program {\it
Mathematica} we have computed all 58,481 scalar interactions of the form
(\ref{eq:0pp-form}) in the level (6, 12) truncation of the theory.
There are 160 scalar fields $\phi_j (p)$ at even levels $\leq 6$, including
longitudinal polarizations of tensor fields, and 31
momentum-independent scalar fields $\phi_i (0)$, each of which takes a
nonzero value in the vacuum $\Phi_0$.  (Actually, the vacuum lies in
a subspace ${\cal H}_1$ of the full scalar field space
\cite{Sen-universality}, but we do not use this decomposition in our
analysis).  As a check on our calculations we have also computed all
the coefficients associated with gauge transformations
(\ref{eq:gauge}) where one of the three fields involved has vanishing
momentum.  We have verified that our terms of the form
(\ref{eq:0pp-form}) give rise to an action (in the perturbative
vacuum) which is invariant at order $g^1$ under arbitrary
momentum-independent gauge transformations and a random sampling of
momentum-dependent gauge transformations.

Using the complete set of terms of the form (\ref{eq:0pp-form}) in the
level (6, 12) truncation, we have calculated all the quadratic terms
for even-level scalars
in the action (\ref{eq:SFT-t-action}) around the perturbatively stable
vacuum.  The details of this action are far too lengthy to appear in
print, but will be made available in the future in electronic form.
In the remainder of this note we summarize the results of using this
quadratic action to study the spectrum of open string states in the
theory (\ref{eq:SFT-t-action}).  Earlier attempts to study the
spectrum of physical states in a subset of the level (2, 6) truncation
appeared in \cite{ks-open,Hata-Teraguchi}.

\section{BRST Cohomology}

The spectrum of physical states in the theory (\ref{eq:SFT-t-action})
is given by the BRST cohomology
\begin{equation}
{\rm Ker}\,\tilde{Q}_1/{\rm Im}\,
\tilde{Q}_0,
\label{eq:}
\end{equation}
where $\tilde{Q}_n$ describes the action of the BRST
operator $\tilde{Q}$ on a string field of ghost number $n$.
{}From (\ref{eq:q2}), it follows that $\tilde{Q}_1 \tilde{Q}_0 = 0$ in
the full string field theory.
The
states associated with vanishing eigenvalues of the kinetic operator
$\tilde{Q}_1$ at a fixed value of $p^2$  are the $\tilde{Q}$-closed states in the
theory.  Two $\tilde{Q}$-closed states are physically equivalent if they differ by
a $\tilde{Q}$-exact state $\tilde{Q}_0\Lambda$ where $\Lambda$ is a
string field 
of ghost number 0.  

Level truncation of the open string field theory  breaks the general
gauge invariance (\ref{eq:gauge}) at order $g^2$, although gauge
invariance is preserved at order $g^0$ and $g^1$.  This breaking of
gauge invariance means that the level-truncated BRST operator no
longer squares to zero.  In other words,
\begin{equation}
\tilde{Q}^{(L, I)}_1 \tilde{Q}^{(L, I)}_0 \neq 0\,
\label{eq:q10n}
\end{equation}
where $\tilde{Q}^{(L, I)}_n$ is the level $(L, I)$ truncated
approximation to $\tilde{Q}_n$.
The inequality (\ref{eq:q10n})
means that $\tilde{Q}$-closed states which are also $\tilde{Q}$-exact in the full string
field theory (\ref{eq:SFT-t-action}) will be approximated in the level
truncated theory by $\tilde{Q}$-closed states which are not precisely $\tilde{Q}$-exact.  This
fact makes the identification of physical states in the theory only
possible in an approximate sense.

We have systematically computed $\tilde{Q}$-closed states in the level-truncated
theory by finding values of $M^2 = -p^2$
where
\begin{equation}
\det \tilde{Q}^{(L, I)}_1= 0
\label{eq:determinant}
\end{equation}
and then computing the eigenvectors associated with the vanishing
eigenvalues.   

As an example of this computation, consider the level (0, 0)
truncation of the theory, which includes only the tachyon field
$\phi$.  The quadratic term for the tachyon field in the nontrivial
vacuum is
\begin{equation}
\phi (-p) \left[ \frac{p^2 -1}{2}  + g \kappa
\left( \frac{16}{27} \right)^{p^2}
\cdot 3 \langle \phi \rangle \right] \phi (p)\,.
\end{equation}
The determinant of $\tilde{Q}^{(0, 0)}_1$ is simply the quantity
in square brackets.  This quantity does not vanish for any real value
of $p^2$, so there are no $\tilde{Q}$-closed states in the spectrum at this level
\cite{ks-open}.

\begin{figure}
\epsfig{file=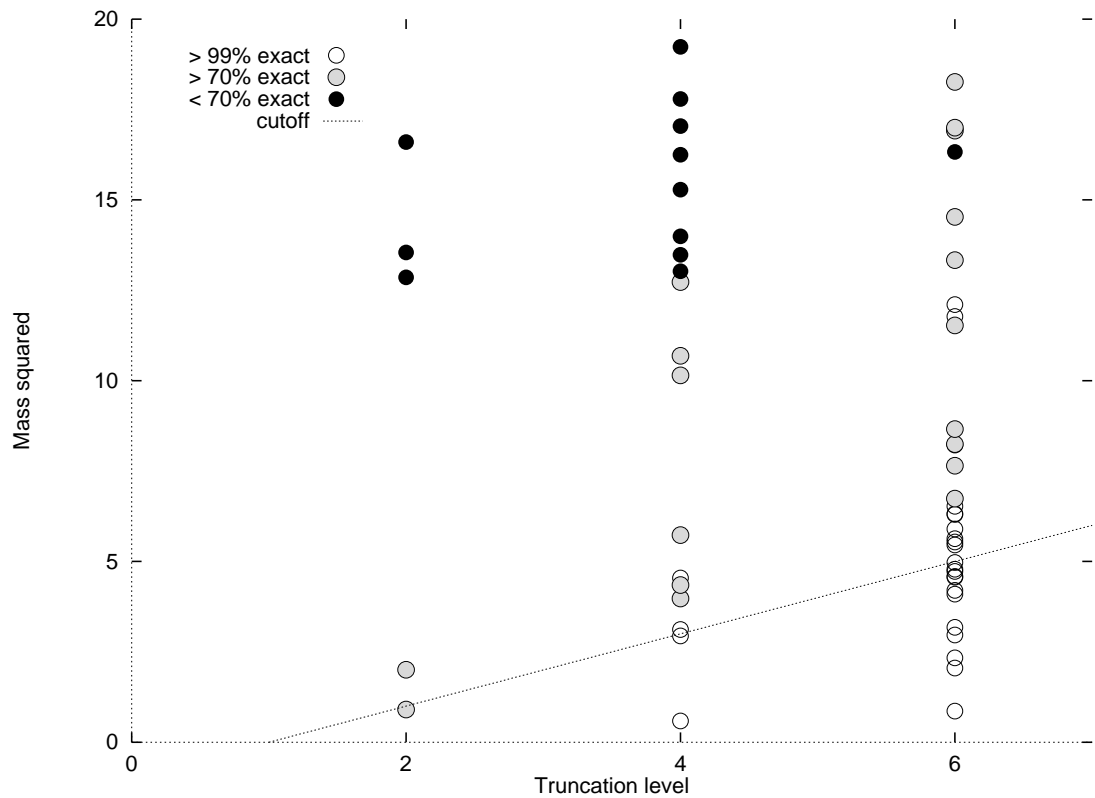,width=15cm}
\caption[x]{\footnotesize Spectrum of $\tilde{Q}$-closed states in level
truncations (0, 0), (2, 6), (4, 12) and (6, 12).  States below the
cutoff $M^2 = L-1$ lie mostly in the exact subspace, confirming Sen's
conjecture}
\label{f:spectrum}
\end{figure}

In the level (2, 6) truncation there are seven scalar fields to be
considered, associated with the Fock space states
\begin{eqnarray}
| 0; p \rangle, &  &  \left( \alpha_{-1} \cdot \alpha_{-1} \right)| 0; p
\rangle, \nonumber\\
\left( \alpha_0 \cdot \alpha_{-2} \right) | 0; p \rangle, &\hspace*{0.5in} & 
\left( \alpha_0 \cdot \alpha_{-1} \right)^2 | 0; p \rangle,
\label{eq:2-states}\\
b_{-1} c_{-1}| 0; p \rangle, &  & 
\left( \alpha_0 \cdot \alpha_{-1} \right) b_{-1} c_{0} | 0; p \rangle,
\nonumber\\ 
b_{-2} c_{-0}| 0; p \rangle & & \nonumber
\end{eqnarray}
At this level of truncation, using the vacuum expectation values
determined in \cite{Moeller-Taylor} with the level (10, 20)
truncation, we found five values of $p^2$ where 
$\det \tilde{Q}^{(2, 6)}_1 = 0$, associated with states having
\begin{equation}
M^2 = 0.9067, \;\;\;2.0032, \;\;\;12.8566, \;\;\;13.5478,
\;\;\;16.5998
\end{equation}
in units where $M^2 = -1$ for the tachyon\footnote{Note: a similar
calculation was done at level (2, 6) in Feynman-Siegel gauge by
Kostelecky and Samuel \cite{ks-open}.  They did not check, however,
whether their spectrum was associated with physical or exact states.
Our calculation can be restricted to this gauge, where it agrees for
the most part (but not exactly) with their calculation.}.  At level
(4, 12) we found 18 $\tilde{Q}$-closed states with $M^2 < 20$, of which the
lightest has $M^2 = 0.58817$.  At level (6, 12) we found 33 $\tilde{Q}$-closed
states with $M^2< 20$, of which the lightest has $M^2= 0.85562$.  The
complete set of $\tilde{Q}$-closed states we found\footnote{Our algorithm for
locating momenta associated with $\tilde{Q}$-closed states proceeded by
calculating the determinant of $\tilde{Q}^{(L, I)}_1$ at equally spaced values
of $p$ (with $\Delta p = 0.0001$) and looking for changes of sign in
the determinant.  The spacing of our $p$ values was significantly less
than the smallest distance we observed between $\tilde{Q}$-closed states (0.0039),
so we believe that we have found all the $\tilde{Q}$-closed states at $M^2 < 20$.
Some possibility remains that we have missed pairs of $\tilde{Q}$-closed states
which are very close in momentum.  It is remotely possible that
physical  states are hiding in such closely spaced pairs of $\tilde{Q}$-closed
states.} is graphed in Figure~\ref{f:spectrum}.

To test the $\tilde{Q}$-exactness of a given $\tilde{Q}$-closed state at level $(L, I)$, we
computed $\tilde{Q}^{(L, I)}_0 \Lambda_i$ for each ghost number zero
field $\Lambda_i$ with level $\leq L$.  The span of the fields
$\tilde{Q}^{(L, I)}_0 \Lambda_i$ gives an approximation to the
subspace of $\tilde{Q}$-exact states at each level.  Suppose that $\{e_i\}$ is an
orthonormal basis for this subspace and $s$ is one of the $\tilde{Q}$-closed
states we found.  We can then measure the extent to which a state is
$\tilde{Q}$-exact by the norm squared of its projection onto the $\tilde{Q}$-exact subspace.
Explicitly,
\begin{equation}
{\rm  fraction\ in\ exact\ subspace} = \frac{\sum_{i} (s \cdot e_i)^2 }{s
\cdot s} \, .
\label{eq:p-exact}
\end{equation}
There is no natural positive definite inner product defined on the
single string Hilbert space ${\cal H}$, so to compute
(\ref{eq:p-exact}) we had to make an ad hoc choice of such an inner
product.  We did the calculation using two choices for this inner
product, and found similar results in both cases.  The first choice,
which seems most natural, is to take the  inner
product $\langle s | s \rangle$ with $p  \rightarrow | p |$
in the matter sector and a Kronecker
delta function in the ghost sector.  The second inner product we tried
was simply defined by a Kronecker delta function on a basis of states
spanned by all possible scalar products of matter and ghost operators
(such as (\ref{eq:2-states}) at level (2, 6), giving a unit
normalization to each of these states).  Using these two definitions
of the inner product, we find for example that the $\tilde{Q}$-closed
state at $M^2 =0.9067$ found in the (2, 6) truncation lies $97.90\%$
in the exact subspace using the first inner product, and $95.24\%$ in
the exact subspace using the second inner product.  In the remainder
of this note all calculations use the first definition of the inner
product.

In the full string field theory, there are continuous families of
$\tilde{Q}$-closed states which are also $\tilde{Q}$-exact at all $p$,
given by states of the form $\tilde{Q}_0| s; p \rangle$.  In the level
truncation approximation we expect these continuous families to be
replaced by a discrete spectrum of almost-exact states, approaching a
continuous distribution as the level of truncation is increased.  
The
extent to which we see a continuous distribution of $\tilde{Q}$-exact
states arising in the level-truncation approximation to the theory
around the vacuum $\Phi_0$ is a measure of how well level truncation
works in the new vacuum, and how close the level-truncation
approximation comes to giving a BRST operator satisfying $\tilde{Q}_1
\tilde{Q}_0 = 0$.  A complete list of $\tilde{Q}$-closed states at
$M^2 < 20$ and the exactness of these states is given in
Table~\ref{t:table}; qualitative results for the exactness of all
$\tilde{Q}$-closed states are depicted in Figure~\ref{f:spectrum}.  As
we would hope, as the level of truncation is increased we see a
discrete distribution of almost-exact states which become both more
exact and more closely spaced as the level of truncation is lifted.
We interpret these almost-exact states as the remnant in the
level-truncated theory of the continuous families of $\tilde{Q}$-exact
states in the full theory.

Physical states in the theory correspond to $\tilde{Q}$-closed states
satisfying $\tilde{Q}_1| s; p \rangle = 0$ which are not
$\tilde{Q}$-exact.  Because states in the cohomology of $\tilde{Q}$
will not be removed by a generic small perturbation, all physical
states in the theory should appear in the level truncation as
$\tilde{Q}$-closed states with $-p^2$ approaching some fixed value
$M^2$ as the level of truncation is increased.  To verify Sen's
conjecture, we would hope to find that the states lying below the
cutoff $M^2 = L-1$ are all approximately $\tilde{Q}$-exact, to the
precision allowed by the level-truncation approximation, so that no
physical states appear in the limiting theory.  Indeed, we find that
beyond the level (2, 6) truncation all states below the cutoff lie
more than $99\%$ in the exact subspace, using either choice of inner
product described above.  For example, the lowest lying states
mentioned above in the level truncations (2, 6), (4, 12) and (6, 12)
lie 97.90\%, 99.990\% and 99.997\% in the exact subspace.  We would
expect physical states in the theory to appear consistently in each
level truncation as states with significant components outside the
exact subspace, since the average state in the level-truncated space
lies less than $35\%$ in the exact space.  We see no sign in our data
of such physical open string states at low levels, even up to values
of $M^2$ several times larger than the cutoff.  We interpret this
result as strong evidence for Sen's conjecture that there are no
physical open string excitations around the vacuum $\Phi_0$, and that
this string field configuration should be identified with the closed
string vacuum.

It may seem surprising that we expect to see the physical states in
the cohomology of $\tilde{Q}$, which form a set of measure zero in the
full space of $\tilde{Q}$-closed states, through this approach.  The
difference between the behavior of physical and exact states under
level truncation of $ \tilde{Q}$ can be understood by considering the
behavior of the zeros of the functions $f (x) = x-1$ and $g (x) = 0$
under a small perturbation by a noise function $\eta (x)$.  In the
first case, generically $f (x) + \eta (x)$ will have a single zero
near $x = 1$.  In the second case, $g (x) + \eta (x)$ will develop a
discrete spectrum of randomly spaced zeros.  The physical states in
the cohomology of $\tilde{Q}$ are controlled by functions like $f
(x)$, while the continuous families of exact states are controlled by
functions like $g (x)$.  While this argument suggests that physical
states should indeed continue to be present in the level-truncation
approximation, as a check on our methodology we have used the same
method we used to compute the approximate cohomology of $\tilde{Q}$ to
compute the approximate cohomology of the BRST operator $Q$ in the
perturbative vacuum, after adding a small random perturbation $\hat{Q}
= Q + \eta$.  We find that unless the perturbation $\eta$ is large
enough to dominate the system ({\it e.g.}, by pushing the exactness of
a generic $\hat{Q}$-closed state below $90\%$), the physical states at
$M^2 = -1$ and $M^2 = 3$ are easily distinguishable in a level (4, 12)
truncation of the theory.
\begin{table}
\begin{center}
\begin{tabular}{|| c | | c || r| r | | c || r | r| | c | | r | r | |}
\hline
\hline
$(L, I)$ &\hspace*{0.05in}& $M^2$&\% exact &\hspace*{0.05in} & $M^2$&\% exact
&\hspace*{0.05in}& $M^2$&\% exact \\
\hline
\hline
& & & & & & & & &\\
\hline
\hline
(2, 6)& & 0.9067 & 97.90\% &  &
2.0032 &93.79\%&
&12.8566 & 64.17\%\\
&& 13.5478 & 5.50\% & & 16.5998 & 2.56\% &  &
&\\
\hline
\hline
& & & & & & & & &\\
\hline
\hline
(4, 12) & &
0.5882 &99.99\% & &
 2.9412 &99.94\% & &
 3.1163 &99.97\% \\
& & 3.9757 &98.51\% & &
 4.3462 &98.92\% & &
 4.5429 &99.07\% \\
& & 5.7318 &98.28\% & &
 10.1466 &80.96\% & &
 10.6907 &98.06\% \\
& &12.7265 &73.42\% & &
 13.0284 &52.71\% & &
13.4834 &37.09\% \\
& & 13.9911 &12.86\% & &
15.2853 &58.25\% & &
 16.2490 &66.20\% \\
& &17.0407 &13.88\% & &
 17.7912 &14.72\% & &
19.2337 &35.80\% \\
\hline
\hline
& & & & & & & & &\\
\hline
\hline
(6, 12) & &
  0.8632& 99.997\% & &
  2.0525& 99.982\% & &
  2.3355& 99.976\% \\
& &  2.9664& 99.997\% & &
  3.1800& 99.998\% & &
  4.0961& 99.999\% \\
& &  4.2023& 99.999\% & &
  4.5645& 99.999\% & &
  4.5869& 99.999\% \\
& &  4.7265& 99.999\% & &
  4.7841& 99.993\% & &
  4.9703& 99.994\% \\
& &  5.4552& 99.984\% & &
  5.5382& 99.976\% & &
  5.6285& 99.992\% \\
& &  5.8999& 99.988\% & &
  6.3008& 99.986\% & &
  6.3204& 99.265\% \\
& &  6.5285& 99.986\% & &
  6.7381& 98.328\% & &
  7.6480& 97.672\% \\
& &  8.2205& 99.936\% & &
  8.2441& 98.748\% & &
  8.6604& 96.683\% \\
& & 11.5289& 98.958\% & &
 11.7778& 99.652\% & &
 12.1027& 99.529\% \\
& & 13.3346& 88.497\% & &
 14.5313& 94.919\% & &
 16.3295& 52.298\% \\
& & 16.9177& 90.786\% & &
 16.9991& 79.305\% & &
 18.2649& 86.453\% \\
\hline
\hline
\end{tabular}
\end{center}
\caption[x]{\footnotesize Masses and exactness of all $\tilde{Q}$-closed states
found in level truncations (2, 6), (4, 12), and (6, 12) with $M^2 < 20$.}
\label{t:table}
\end{table}

\section{Discussion}

We have explicitly calculated the quadratic terms in the open string
field theory action around the nonperturbative vacuum $\Phi_0$ in the
(6, 12) level truncation.  We computed the BRST cohomology by
computing all closed states under the truncated BRST operator
$\tilde{Q}^{(L, I)}_1$, and comparing with the subspace of exact
states formed by the operator $\tilde{Q}^{(L, I)}_0$.  We found
evidence that all $\tilde{Q}$-closed states in the theory become
$\tilde{Q}$-exact in the limit when fields of all levels are included.

There are several directions in which it would be interesting to
proceed, given the results in this letter.  For one thing, it would be
very nice to have a better conceptual understanding of the decoupling
of open string states in the perturbatively stable vacuum.  While some
interesting perspectives on this phenomenon have been given
\cite{Yi-membranes,bhy,WT-mass,Sen-fundamental,Gerasimov-Shatashvili,Gerasimov-Shatashvili-2,kls,Chalmers},
a convincing picture which explains the classical decoupling of open
strings in the cubic string field theory picture has yet to be given.
An intriguing suggestion for the form of the string field theory in
the nontrivial vacuum was made in \cite{rss,rss2}, where it was
suggested that the new BRST operator $\tilde{Q}$ can be related
through a field definition to a pure ghost operator such as $c_0$ or
more generally $ \sum a_n (c_n + (-1)^nc_{-n})$.  It would be very
interesting to use the explicit form of $\tilde{Q}$ which we have
computed in level truncation to prove or disprove this conjecture.
Finally, a question of fundamental importance is to what extent the
cubic open string field theory in the perturbatively stable vacuum
contains closed string excitations.  Sen's conjectures suggest that
it might be possible to give a direct description of asymptotic closed
string states in terms of the open string field theory degrees of
freedom in the vacuum $\Phi_0$.  If such a description could be made
explicit, it would lead to new insight into the nature of closed
string field theory and the structure of D-branes as closed string
solitons.

\section*{Acknowledgments}

We would like to thank D.\ Gross, N.\ Moeller, J.\ Polchinski, A.\ Sen
and B.\ Zwiebach for helpful discussions in the course of this work.
Particular thanks to A.\ Sen and B.\ Zwiebach for constructive
comments on an early version of this manuscript.
WT would like to thank the Institute for Theoretical Physics in Santa
Barbara for hospitality during the latter part of this work.  The work
of IE was supported in part by a National Science Foundation Graduate
Fellowship and in part by the DOE through contract
\#DE-FC02-94ER40818.  The work of WT was supported in part by the A.\
P.\ Sloan Foundation and in part by the DOE through contract
\#DE-FC02-94ER40818.

\bibliographystyle{plain}

\end{document}